\documentclass[useAMS,usenatbib]{mn2e}
\usepackage{graphicx}
\usepackage{times}

\newcommand{\kms}{\mbox{$\mathrm{km\,s^{-1}}$}}
\newcommand{\MSUN}{\mbox{$\mathrm{M_{\odot}}$}}
\newcommand{\RSUN}{\mbox{$\mathrm{R_{\odot}}$}}
\newcommand{\targ}{TYC\,6760-497-1}
\newcommand{\gae}{\lower 2pt \hbox{$\, \buildrel{\scriptstyle >}\over
    {\scriptstyle \sim}\,$}}

\title[The close binary TYC 6760-497-1]{The first pre-supersoft X-ray binary}

\author[S. G. Parsons et al.]{S.~G.~Parsons$^{1}$\thanks{steven.parsons@uv.cl},
M.~R.~Schreiber$^{1,2}$,
B.~T.~G{\"a}nsicke$^{3}$,
A.~Rebassa-Mansergas$^{4}$,
\newauthor
R.~Brahm$^{5,6}$,
M.~Zorotovic$^{1}$,
O.~Toloza$^{3}$,
A.~F.~Pala$^{3}$,
C.~Tappert$^{1}$,
A.~Bayo$^{1,2}$
\newauthor
and A.~Jord{\'a}n$^{5,6}$
\\
$^{1}$ Instituto de F{\'i}sica y Astronom{\'i}a, Universidad de
Valpara{\'i}so, Avenida Gran Bretana 1111, Valpara{\'i}so, Chile\\
$^{2}$ Millenium Nucleus "Protoplanetary Disks in ALMA Early Science",
Universidad de Valparaiso, Valparaiso 2360102, Chile\\
$^{3}$ Department of Physics, University of Warwick, Coventry CV4 7AL, UK\\
$^{4}$ Kavli Institute for Astronomy and Astrophysics, Peking University,
Beijing 100871, China\\
$^{5}$ Instituto de Astrofísica, Facultad de Física, Pontificia Universidad
Cat{\'o}lica de Chile, Av. Vicu{\~n}a Mackenna 4860, 7820436 Macul, Santiago,
Chile\\
$^{6}$ Millennium Institute of Astrophysics, Av. Vicu{\~n}a Mackenna 4860,
7820436 Macul, Santiago, Chile}
\begin{document}
\input{references.cls}
\date{Accepted 2015 June 22.  Received 2015 June 15; in original form 2015 March 25}

\pagerange{\pageref{firstpage}--\pageref{lastpage}} \pubyear{2015}

\maketitle

\label{firstpage}

\begin{abstract}

We report the discovery of an extremely close white dwarf plus F dwarf
main-sequence star in a 12 hour binary identified by combining data from the
RAdial Velocity Experiment (RAVE) survey and the Galaxy Evolution Explorer 
(GALEX) survey. A combination of spectral energy distribution fitting
and optical and Hubble Space Telescope ultraviolet spectroscopy allowed us to
place fairly precise constraints on the physical parameters of the binary. The
system, \targ, consists of a hot $T_\mathrm{eff}\sim20,000K$,
$M_\mathrm{WD}\sim0.6\MSUN$ white dwarf and an F8 star 
($M_\mathrm{MS}\sim1.23\MSUN$, $R_\mathrm{MS}\sim1.3\RSUN$) seen at
a low inclination ($i\sim37^\circ$). The system is likely the descendent of a
binary that contained the F star and a $\sim$2{\MSUN} A-type star that filled
its Roche-lobe on the thermally pulsating asymptotic giant branch,
initiating a common envelope phase. The F star is extremely close to
Roche-lobe filling and there is likely to be a short phase of thermal
timescale mass-transfer onto the white dwarf during which stable hydrogen
burning occurs. During this phase it will grow in mass by up to 20 per
cent, until the mass ratio reaches close to unity, at which point it will
appear as a standard cataclysmic variable star. Therefore, {\targ} is the
first known progenitor of a super-soft source system, but will not undergo a
supernova Ia explosion. Once an accurate distance to the system is determined
by Gaia, we will be able to place very tight constraints on the stellar and
binary parameters.

\end{abstract}

\begin{keywords}
binaries: close -- stars: white dwarfs -- stars: early-type -- stars: evolution
\end{keywords}

\section{Introduction}

The unique properties of Type Ia Supernovae (SN\,Ia) as distance indicators,
sufficiently bright to serve as yardsticks on cosmological scales
\citep[e.g.][]{branch92}, has resulted in them becoming some of the
most important objects in the Universe, having led to the discovery of its
accelerating expansion \citep{riess98, perlmutter99}.

Although it is generally accepted that SN\,Ia are related to
the thermonuclear ignition of a white dwarf that surpassed the Chandrasekhar
mass limit, there is no consensus yet on the pathways leading to
the explosion. The two main formation channels are thought to be the
``single-degenerate'' channel, where a white dwarf accretes material from a 
non-degenerate companion via Roche-lobe overflow \citep{whelan73} during a
phase of thermal time-scale mass transfer known also as the super soft source
(SSS) phase, and the ``double-degenerate'' channel in which the explosion is
the result of the merger of two white dwarfs \citep{webbink84}. In recent
years the idea of a ``double-detonation'' scenario has also been advanced. In
this case a layer of helium on the surface of the white dwarf detonates,
triggering the explosion of the underlying core either directly, or via a
compressional shock wave \citep{fink07,fink10,shen14}, hence it is possiblevia this channel to detonate the white dwarf while its mass is still below the
Chandrasekhar limit. In theory these ``sub-Chandrasekhar SN\,Ia'' are possible
via both main channels. In the single degenerate channel the white dwarf
accretes helium-rich material from a donor star, building up a large surface
layer \citep{tutukov96}, whilst the merger of a carbon-oxygen core white dwarf
with a low-mass helium core white dwarf can lead to helium ignition and hence a
supernova via the double degenerate channel \citep{fink07}.

The search for Galactic SN\,Ia progenitors from either of the channels is a
difficult task. In the single degenerate case very few SSS systems
sufficiently nearby for detailed parameter studies are known
\citep{greiner00}. This is due to a combination of the short duration of the
SSS phase and because their very soft X-ray emission is easily absorbed by
neutral hydrogen in the galactic plane. Whilst double white dwarf binaries are
intrinsically faint objects and difficult to distinguish from single white
dwarfs without dedicated spectroscopic monitoring \citep[see][for
example]{napiwotzki03}. However, in both cases the
progenitor systems are the descendents of detached white dwarf plus F, G or
early K type main-sequence star companions. With no ongoing accretion
these systems are easy to characterise. Furthermore, they are expected to be
numerous \citep{holberg13} and so are optimal for population studies and hence
testing both SN\,Ia progenitor channels.

However, identifying white dwarfs with early type companions has been
extremely difficult until recently. This is due to the fact that the main
sequence star completely outshines the white dwarf at optical wavelengths
\citep{rebassa12b}, hence a combination of optical and ultraviolet (UV)
coverage is required to find these systems. \citet{maxted09} attempted to
detect systems of this type by combining data from the Galaxy Evolution
Explorer (GALEX) survey and the Sloan Digital Sky Survey (SDSS), but were
limited by the optical colour selection of their main-sequence stars. Whilst
\citet{burleigh97} found only four unresolved white dwarf plus FGK
main-sequence star systems searching for extreme-UV and soft X-ray excesses.

We have recently begun a project combining the large dataset of the RAdial
Velocity Experiment (RAVE) survey \citep{kordopatis13} in the southern
hemisphere and the LAMOST (Large Sky Area Multi-Object Fiber Spectroscopic
Telescope) survey in the northern hemisphere, with UV data from the GALEX
survey in order to identify main-sequence F, G and early K stars with
significant UV excesses. 

The RAVE survey spectroscopically observed more than 400,000 bright
(8$<I<$12\,mag) southern hemisphere stars in the spectral region
8410-8794{\AA} (the infrared calcium triplet) with a resolution of
R$\sim$7000. The data were used to establish the basic parameters of each star
(effective temperatures, surface gravities and metallicities) as well as
line-of-sight velocities. We used these data, in conjunction with PHOENIX
synthetic spectra \citep{husser13}, to estimate the UV colours of these stars.
We selected those showing UV excess flux for radial velocity follow-up
observations to detect close binarity, as well as Hubble Space Telescope UV
spectroscopy to confirm the presence of a white dwarf. Full details of this
process and the list of UV-excess objects will be published in a forthcoming
paper (Parsons et~al. in prep.)

As well as providing a crucial test of the single degenerate channel for
SN\,Ia, these detached white dwarf plus early-type main-sequence star binaries
also provide useful constraints for the common envelope phase of binary
evolution, which they passed through when the progenitor of the
white dwarf evolved off the main-sequence. Essentially all current
observational constraints on the common envelope phase are based on systems
containing white dwarfs with low-mass M dwarf companions
\citep[e.g.][]{zorotovic10}. However, the handful of systems with more massive
main-sequence star components already hint that additional energy may
be needed to expel the envelope in these cases \citep{zorotovic14}.

In this paper we present data for the first of our UV-excess objects from the
RAVE sample to be confirmed as a close white dwarf plus main-sequence
binary. The object, {\targ}, is a late-F star. We derive clear observational
constraints on the stellar and binary parameters of the systems, reconstruct
its evolutionary history, predict its future and discuss the implications for
our understanding of close compact binary star evolution. 

\section{Observations and their reduction}

\subsection{Optical echelle spectroscopy}

\subsubsection{FEROS and the Du Pont echelle}

We obtained high resolution spectroscopy of {\targ} with the echelle
spectrograph (R$\sim$40,000) on the 2.5-m Du Pont telescope located at Las
Campanas Observatory, Chile and with FEROS (R$\sim$48,000) on the 2.2-m
Telescope at La Silla, Chile. FEROS covers the wavelength range from
$\sim$3500{\AA} to $\sim$9200\AA. The observations formed part of a program
to detect binarity among a large number of potential white dwarf plus
main-sequence binaries. However, it was immediately clear that {\targ} had a
very short period when we observed a very large velocity shift between our
first two measurements. Therefore, we obtained several spectra of this target
per night throughout the rest of the observing run with the aim of measuring
its period. 

Data obtained with the FEROS and duPont spectrographs were extracted and
analysed with an automated pipeline developed to process spectra coming from
different instruments in an homogeneous and robust manner (\citealt{jordan14},
Brahm et~al in prep.). After performing typical image reductions, spectra
were optimally extracted following \citet{marsh89} and calibrated in wavelength
using reference ThAr Lamps. For FEROS data, the instrumental drift in
wavelength through the night was corrected with a secondary fiber observing a
ThAr lamp. In the case of the duPont data, ThAr spectra were acquired before
and after each science observation. Wavelength solutions were shifted to the
barycenter of the solar system. 

\begin{table}
 \centering
  \caption{Radial velocity measurements for the main-sequence star in \targ.}
  \label{tab:vels}
  \begin{tabular}{@{}lccc@{}}
  \hline
  BJD             & Velocity & Uncertainty & Telescope/      \\
  (mid-exposure)  & (\kms)   & (\kms)      & instrument      \\
  \hline
 2456811.5672390 &  30.33  & 0.50       & Du Pont/Echelle \\
 2456827.6293689 & -48.79  & 0.48       & MPG2.2/FEROS    \\
 2456828.4841811 &  53.36  & 0.28       & MPG2.2/FEROS    \\
 2456828.5729529 & -13.12  & 0.28       & MPG2.2/FEROS    \\
 2456828.5904822 & -26.22  & 0.28       & MPG2.2/FEROS    \\
 2456828.6012976 & -34.38  & 0.27       & MPG2.2/FEROS    \\
 2456828.6663049 & -61.48  & 0.30       & MPG2.2/FEROS    \\
 2456828.6775301 & -63.33  & 0.30       & MPG2.2/FEROS    \\
 2456828.7213267 & -55.85  & 0.28       & MPG2.2/FEROS    \\
 2456829.5163187 &  32.99  & 0.29       & MPG2.2/FEROS    \\
 2456829.5491571 &   7.39  & 0.27       & MPG2.2/FEROS    \\
 2456829.5939247 & -29.35  & 0.30       & MPG2.2/FEROS    \\
 2456829.6737526 & -62.99  & 0.31       & MPG2.2/FEROS    \\
 2456829.6855582 & -61.90  & 0.29       & MPG2.2/FEROS    \\
 2456829.6963120 & -60.80  & 0.29       & MPG2.2/FEROS    \\
 2456829.7123239 & -58.27  & 0.29       & MPG2.2/FEROS    \\
 2456829.7603942 & -33.58  & 0.30       & MPG2.2/FEROS    \\
 2456829.7718605 & -24.86  & 0.29       & MPG2.2/FEROS    \\
 2456829.7818813 & -15.99  & 0.28       & MPG2.2/FEROS    \\
 2456834.6122883 & -49.91  & 0.30       & MPG2.2/FEROS    \\
 2456834.6226973 & -52.15  & 0.29       & MPG2.2/FEROS    \\
 2456835.7338779 & -41.03  & 0.29       & MPG2.2/FEROS    \\
  \hline
  \end{tabular}
\end{table}

\subsubsection{X-shooter}

A medium resolution spectrum (R$\sim$5,000) of {\targ} was obtained with
X-shooter \citep{dodorico06} mounted at the Cassegrain focus of VLT-UT2 at
Paranal on the 26th of February 2015. X-shooter is comprised of three
detectors that enable one to obtain simultaneous data from the UV cutoff at
0.3$\mu$m to the $K$-band at 2.4$\mu$m. Our data consisted of three 60 second
exposures, which were reduced using the standard pipeline release of the
X-shooter Common Pipeline Library (CPL) recipes (version 2.5.2). The
instrumental response was removed and the spectrum flux calibrated by
observing the spectrophotometric standard star CD-38\,10980 and dividing it by
a flux table of the same star to produce the response function. The
spectrum was extinction corrected but not corrected for telluric features.

\subsection{Ultraviolet spectroscopy}

{\targ} was observed with the Hubble Space Telescope (HST) on the 9th of
January 2015 with the Space Telescope Imaging Spectrograph (STIS) as part of
program GO 13704. We used the G140L grating centered on 1425{\AA}, for one
spacecraft orbit, resulting in a total exposure time of 2381 seconds. The data
were processed using {\sc calstis} V3.4. We de-reddened the STIS spectrum
  using a value of $E(B-V)=0.103$, determined from our SED fit to the
  main-sequence star (see Section~\ref{sec:sed}).

\section{Results}

\subsection{Optical spectra}

\begin{figure}
\begin{center}
 \includegraphics[width=\columnwidth]{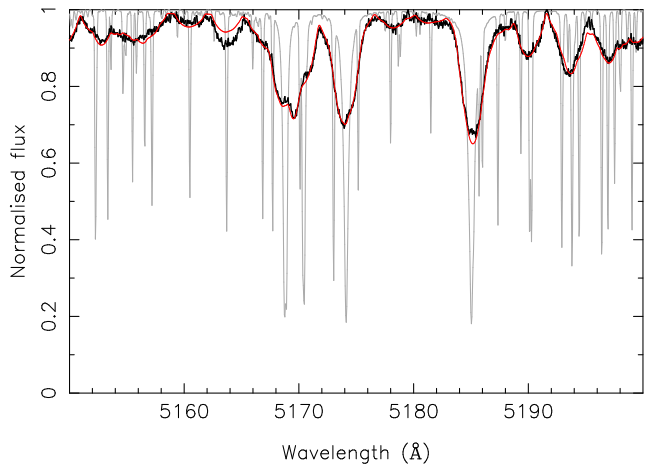}
 \caption{Observed FEROS spectrum (black line) of {\targ} with the best fit
   model overplotted (red line, $v_\mathrm{rot}\sin{i}=75$\kms). We also plot
   the model spectra without any rotational broadening (grey line) to
   highlight the rapid rotation of the star.}
 \label{fig:specfit}
 \end{center}
\end{figure}

Radial velocities were computed from our optical echelle spectra using the
cross-correlation technique against a binary mask representative of a G2-type
star. The uncertainties in radial velocity were computed using scaling
relations \citep[for more detail see][]{jordan14} with the signal-to-noise
ratio and width of the cross-correlation peak, which were calibrated with
Monte Carlo simulations. The results are listed in Table~\ref{tab:vels}.
  We note however, that these relations were calibrated using slowly rotating
  stars and hence the uncertainties may be underestimated for rapidly rotating
  stars, such as the one in {\targ}.

We also estimated the projected rotational velocity ($v_\mathrm{rot}\sin{i}$)
of the main-sequence star in {\targ} by comparing the observed FEROS spectra
against a synthetic grid of stellar spectra \citep{coelho05}. The synthetic
spectra were degraded to the resolution of FEROS by convolving them with a
Gaussian with $R = \lambda / \delta \lambda = 53000$ and then further degraded
to different values of $v_\mathrm{rot}\sin{i}$ using the limb-darkening
coefficients of \citet{claret04}. The optimal fit (which also yields a set of
stellar parameters) was found by chi-square minimisation in 3 echelle order of
the spectra which include the zone of the magnesium triplet (5000-5500\AA). The
measured rotational broadening was $v_\mathrm{rot}\sin{i}=75\pm3$\,\kms. We
also obtained the following stellar parameters:
$T_\mathrm{eff}=5750\pm200$\,K, $\log{g}=3.75\pm0.3$ (in cgs units),
[Fe/H]=$-0.5\pm0.2$. Figure~\ref{fig:specfit}  shows the observed spectra with
the best fitted model, as well as the same model with no rotational
broadening. This high rotation rate means that the fitted stellar parameters
from the spectral fit are not necessarily reliable since most of the weak
lines that have a $\log{g}$ dependence are completely smeared out. Moreover,
the system is extremely close to Roche-lobe filling (see
Section~\ref{sec:params}) and hence the effects of gravity darkening and
Roche-distortion likely have an effect on the spectral fit.

\subsection{Orbital period}

\begin{figure}
\begin{center}
 \includegraphics[width=\columnwidth]{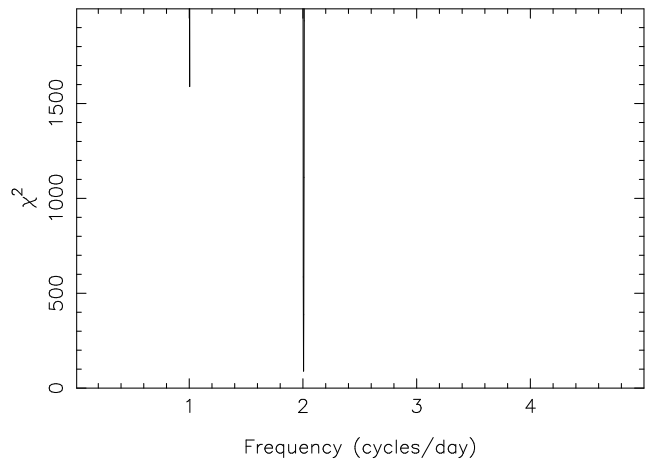}
 \caption{$\chi^2$ plotted against orbital period for the fit to the radial
   velocity measurements. There is a clear minimum at a period of 0.5 days}
 \label{fig:pgram}
 \end{center}
\end{figure}

\begin{figure}
\begin{center}
 \includegraphics[width=\columnwidth]{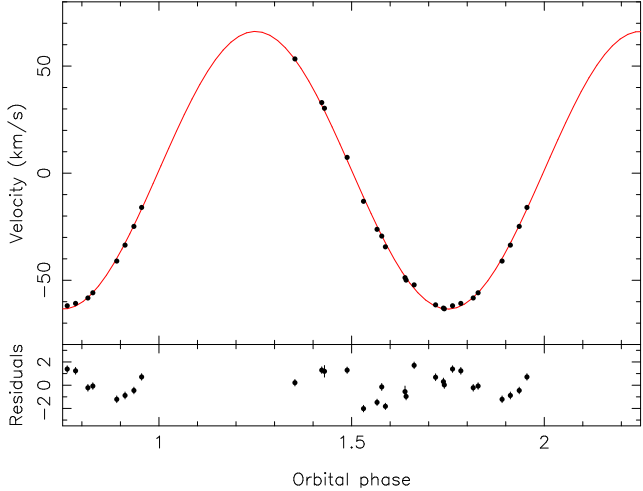}
 \caption{Phase-folded radial velocity plot for the main-sequence star in
   \targ. The lower panel shows the residuals to the best fit.}
 \label{fig:rvplot}
 \end{center}
\end{figure}

We measured the binary period of {\targ} by fitting a constant plus sine wave
to the velocity measurements over a range of periods and computing the
$\chi^2$ of the resulting fit. The result of this is shown in
Figure~\ref{fig:pgram} and shows a clear minimum in $\chi^2$ at a period of
0.5 days. We determine the ephemeris of {\targ} as
\begin{eqnarray}
\mathrm{BJD} = 2456823.81993(40) + 0.498688(26)E,
\end{eqnarray}
where $E$ is the binary phase and phase 0 occurs at the conjunction of the
main-sequence star.

The phase-folded radial velocity plot is shown in Figure~\ref{fig:rvplot}. We
find a radial velocity semi-amplitude for the main-sequence star of
$65.0\pm0.3$\kms and a systemic velocity of $1.4\pm0.2$\kms. The lower panel
of Figure~\ref{fig:rvplot} shows the residuals to the fit, which show
variations larger than their uncertainties. This is likely due to the
main-sequence star's large rotational broadening causing small systematic
errors during the cross correlation process.

\section{System parameters} \label{sec:params}

In this section we detail how we constrained the physical parameters of the
binary.

\subsection{SED fit} \label{sec:sed}

In order to get an initial estimate of the physical parameters of the
main-sequence star in {\targ} and the reddening towards the system
we fitted its spectral energy distribution (SED) using the virtual observatory
SED analyzer (VOSA; \citealt{bayo08}). We convolved our flux calibrated
  X-shooter spectrum with a number of generic narrow and broad band filters
  including Str{\"o}mgren $uvby$, Bessell $BVRI$ Johnson $BVR$ and Cousins
  $RI$ filters. We combined these with archival data from the GALEX
  (\citealt{martin05}), 2MASS \citep{skrutskie06} and WISE \citep{wright10}
  catalogues. We excluded the GALEX data when fitting the SED since we
expected the white dwarf to contribute a non-negligible amount of flux at
these wavelengths. 

\begin{figure}
\begin{center}
 \includegraphics[width=\columnwidth]{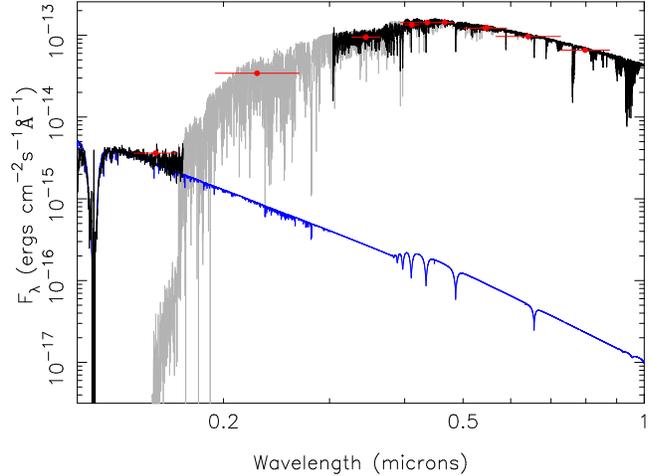}
 \caption{The spectral energy distribution of {\targ} in the UV and optical
   wavelength ranges (the SED fit also included infrared data, but we limit
   the plot to this range to demonstrate the relative contributions of the two
   stars). The black lines are the observed UV HST/STIS spectrum and optical
   X-shooter spectrum (no telluric correction was applied). Broad band
   photometry is shown in red (their errors are too small to see on this
   scale), the best fit BTSettl main-sequence star model is shown in grey and
   was rotationally broadened to match the measured value for the
   main-sequence star in \targ. We also plot in dark blue a TLUSTY/SYNSPEC
   model for a $T_{\mathrm{eff}}=20400$\,K, $\log{g}=8$ white dwarf. The
     fit to the UV part of the spectrum is better illustrated in
     Figure~\ref{fig:stisfit}.}
 \label{fig:sed}
 \end{center}
\end{figure}

We used a large grid of BTSettl \citep{allard12} models to determine the
main-sequence star's parameters and allowed the extinction to vary from
  zero to three times the maximum expected value of A$_\mathrm{V}=0.36$
  \citep{schlafly11}. As expected the fit was insensitive to the metallicity
  of the system and only mildly sensitive to the surface gravity, preferring a
  value of $\log{g}\sim4.0$. However, we found that the temperature was well
  constrained to $T_\mathrm{eff}=6400\pm100$\,K and the extinction to be
  A$_\mathrm{V}=0.32$, implying $E(B-V)=0.103$, in good agreement with what is
  expected from the \citet{schlafly11} maps. The uncertainties were
  determined via a Bayes analysis \citep[for more details see][]{bayo08}.
These measurements are consistent with those from the RAVE survey, in which
{\targ} was determined to have $T_\mathrm{eff}=6150\pm100$\,K and
$\log{g}=3.6\pm0.2$ \citep{kordopatis13}, although these measurements are
likely affected by the large rotational broadening, similar to our FEROS
fit. In Figure~\ref{fig:sed} we show part of the SED in the UV and optical
wavelength ranges with the best fit model and observed spectra
overplotted. While the GALEX NUV measurement shows little evidence of an
excess, the FUV measurement is clearly dominated by the white dwarf.

\subsection{Rotational broadening constraint}

Assuming that the main-sequence star is tidally locked to the white dwarf,
which it should be since the tidal synchronisation timescale for this system is less than 0.5\,Myr \citep{zahn77}, much shorter than the white dwarf's
cooling age (see Section~\ref{sec:past}), the rotational broadening
measurement can be used to place some constraints on the binary parameters via 
\begin{eqnarray}
v_\mathrm{rot}\sin{i} = K_\mathrm{MS} (1+q) \frac{R_\mathrm{MS}}{a},
\end{eqnarray}
where $K_\mathrm{MS}$ is the radial velocity semi-amplitude of the
main-sequence  star,
$q=M_\mathrm{MS}/M_\mathrm{WD}$, the mass ratio of the binary and
$R_\mathrm{MS}/a$ is the radius of the main-sequence star scaled by the orbital
separation. Combining this with Kepler's third law,
\begin{eqnarray}
a^3 = \frac{P^2G M_\mathrm{MS}}{4 \pi^2} \left( 1 + \frac{1}{q} \right),
\end{eqnarray}
where $P$ is the orbital period and $G$ is the gravitational constant, we can
effectively solve for $q$ (hence $M_\mathrm{WD}$) by assuming a mass and
radius for the main-sequence star. Once this is known all the other binary
parameters can then be determined. For example, the orbital inclination via
\begin{eqnarray}
\sin^3i = \frac{P K_\mathrm{MS}^2 (1+q)^2}{2 \pi G M_\mathrm{WD}}.
\end{eqnarray}

\begin{figure}
\begin{center}
 \includegraphics[width=\columnwidth]{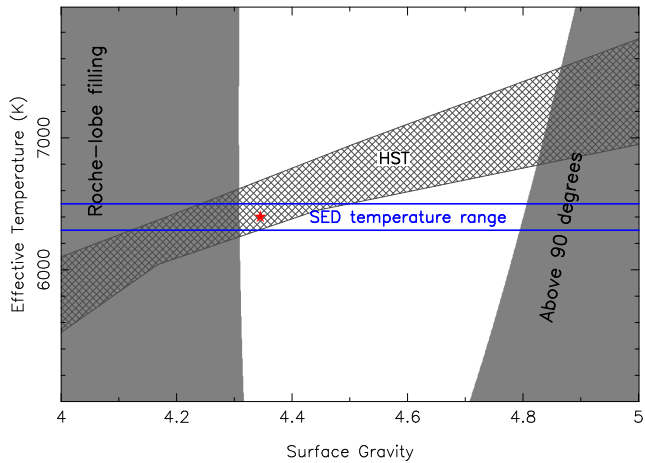}
 \caption{Limits on the physical parameters of the main-sequence star. The
   measured rotational broadening, period and radial velocity amplitude
   exclude the light grey areas as at low surface gravities the star fills its
   Roche lobe, whilst high surface gravities lead to unphysical results. The
   horizonal blue lines indicate the possible temperature range from the SED
   fit. The dark grey hatched area indicates the parameter range that is
   consistent with the fit to the white dwarf's spectrum (see
   Section~\ref{sec:wdfit}). The small region that is consistent with all our
   constraints is highlighted by a red star symbol.} 
 \label{fig:lims}
 \end{center}
\end{figure}

Ideally the fit to the FEROS spectra would yield the main-sequence star's
effective temperature, surface gravity and metallicity, which can then be used
to estimate its mass and radius via the Torres relation \citep{Torres10} and
hence determine the binary parameters. Unfortunately, due to its rapid rotation 
the FEROS fit does not give reliable stellar parameters. Therefore, keeping
the metallicity fixed at the solar value, we used a grid of $T_\mathrm{eff}$,
$\log{g}$ values and the Torres relation to determine the range of possible
binary parameters. The result of this is shown in Figure~\ref{fig:lims}. We
found that at surface gravities below $\sim$4.3 the binary parameters were
such that the main-sequence star should be Roche-lobe filling. Since there is
no evidence for this, we can exclude this region. Furthermore, at very high
surface gravities ($>$4.8), a solution to the binary parameters is not
possible since it would require the inclination to be larger than 90
degrees. Hence this region can also be excluded. However, this still results
in a large range of possible parameters, the white dwarf mass can be anything
between 0.3{\MSUN} and 0.7{\MSUN}, the main-sequence star mass between
1.08{\MSUN} and 1.26{\MSUN} and the inclination larger than 32
degrees. Varying the metallicity does not have a large effect on these
limits.

\begin{figure}
\begin{center}
 \includegraphics[width=\columnwidth]{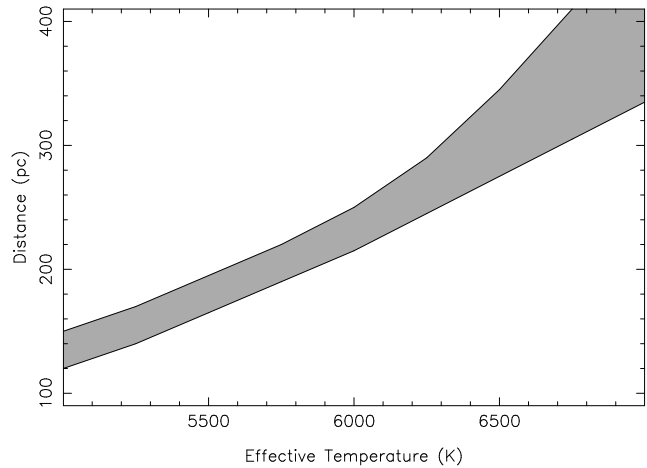}
 \caption{Distance to the main-sequence star in {\targ} as a function of its
   effective temperature. The grey region shows the permitted range.}
 \label{fig:msdist}
 \end{center}
\end{figure}

We estimated the distance to {\targ} using isochrones from the PAdova and
tRieste Stellar Evolution Code (PARSEC) \citep{bressan12}. For a given
temperature we used the isochrones to calculate the absolute magnitude of the
main-sequence star over the full range of allowed surface gravities in the
$BVRIJHK$ bands and compared these to the measured values to calculate the
distance. This results in a range of distances for a given main-sequence star
temperature. We take into account the variations in the calculated
distances from the different bands and the effects of the unknown age of the
star. The result of this is shown in Figure~\ref{fig:msdist}. While the
  unknown age of the star generally has little effect on the predicted
  distance, if it is very young ($<$30\,Myr, which is in fact ruled out by the
  cooling age of the white dwarf) or very old ($>$3.2\,Gyr) the resultant
  distance is somewhat larger than whilst on the main-sequence.

\subsection{White dwarf spectral fit} \label{sec:wdfit}

\begin{figure}
\begin{center}
 \includegraphics[width=\columnwidth]{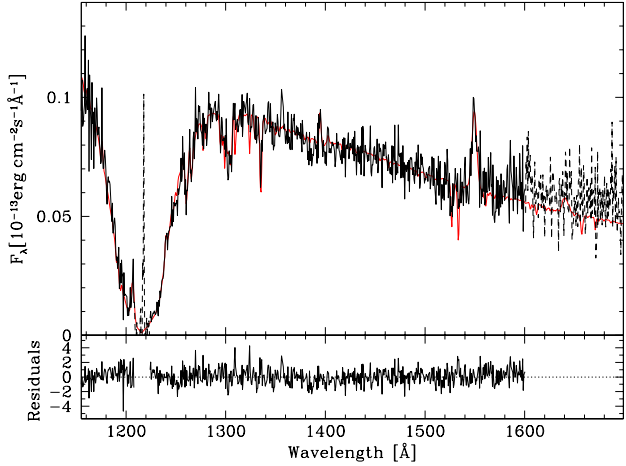}
 \caption{HST/STIS spectrum of the white dwarf in {\targ} with a model fit
   overplotted (red line, $\log{g}=8.0$). The dashed parts were not included
   in the fit, these include the core of the Ly$\alpha$ line, which is
   contaminated by geocoronal emission, and the red end of the spectrum, which
   has low signal-to-noise. There are also several emission lines including the
   quite strong C{\sc iv} 1550{\AA} line. These are likely due to the white
   dwarf capturing material from the wind of the F star.}
 \label{fig:stisfit}
 \end{center}
\end{figure}

We fitted the HST/STIS spectrum of the white dwarf in {\targ} with a grid of
white dwarf models computed using TLUSTY/SYNSPEC \citep{hubeny95}, stepping in
surface gravity from 7.0 to 9.5 in steps of 0.1 (see
Figure~\ref{fig:stisfit}). At each step the best fit to the spectrum gives the
temperature and hence the mass and radius using the cooling models of
\citet{fontaine01}. Scaling the flux then also gives an estimate of the
distance. Hence we are able to determine the white dwarf mass, effective
temperature and distance as a function of the surface gravity. These relations
are illustrated in Figure~\ref{fig:wdlims}. $T_\mathrm{eff}$ and
  $\log{g}$ are correlated parameters, so increasing $\log{g}$ results in a
  higher best-fit $T_\mathrm{eff}$. At the same time, increasing $\log{g}$
  implies a higher mass, and, via the mass-radius relation, a smaller
  radius. This drop in radius dominates over the increase in $T_\mathrm{eff}$
  in the UV flux emitted by the white dwarf, i.e. increasing the surface
  gravity leads to a lower best-fit distance.

Several narrow absorption features are seen in the STIS spectrum,
primarily from Si and C as a result of the white dwarf accreting material from
the wind of the main-sequence star as is commonly seen in other close white
dwarf binaries \citep{tappert11,parsons12,pyrzas12,ribeiro13}. We found
setting the metal abundances in the model to 0.1 times solar for all elements
fits these well.

\subsection{Combined constraints}

\begin{figure}
\begin{center}
 \includegraphics[width=\columnwidth]{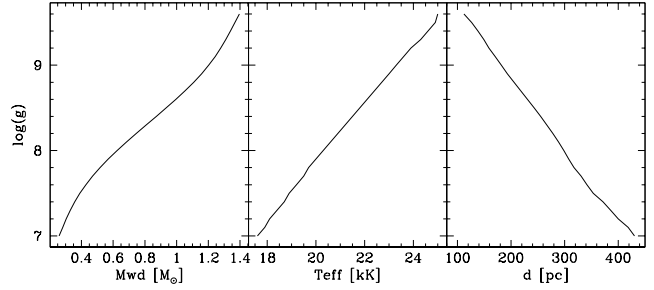}
 \caption{Constraints on the white dwarf's parameters as a function of its
   surface gravity.}
 \label{fig:wdlims}
 \end{center}
\end{figure}

We can combine the constraints from fitting the white dwarf's spectrum with
those from the main-sequence star to determine which set of parameters are
consistent with all our data. For a given main-sequence star effective
temperature, we have a range of possible distances from the isochrone fitting
(see Figure~\ref{fig:msdist}) , which then leads to a range of possible white
dwarf masses from the relations shown in Figure~\ref{fig:wdlims}. The
rotational broadening measurement also yields a white dwarf mass for a given
main-sequence star temperature and gravity, allowing us to determine the range
over which these two measurements are consistent. This range is shown in
Figure~\ref{fig:lims} as the dark grey hatched region. We can instantly
exclude a main-sequence star with a temperature less than $\sim$6200\,K, since
a star this cool would be at a short distance and hence would have a white
dwarf with a mass$>$0.7\MSUN, which is only possible if the system fills its
Roche lobe. It also means that the system must be at a minimum distance of
250\,pc. 

These constraints are also consistent with those from the main-sequence star
temperature determined from the SED fitting ($T_\mathrm{eff}=6400\pm100$\,K),
which is also illustrated in Figure~\ref{fig:lims}. There is therefore a small
region in which all of our measurements and fits are consistent, highlighted
by a red star in Figure~\ref{fig:lims}. The full set of ranges for all the
parameters are given in Table~\ref{tab:params}. 

\section{Discussion}

Having determined the stellar and binary parameters of {\targ}, we can now
investigate the evolution of the system and its implications for models of
compact binary star evolution and SN\,Ia formation channels. We start by
reconstructing the systems evolutionary history.

\subsection{Constraints on common envelope evolution} \label{sec:past}

Common envelope evolution is typically described with a parametrized energy
equation: 
\begin{equation}\label{eq:alpha}
E_\mathrm{bind} = \alpha_{\mathrm{CE}}\Delta E_\mathrm{orb},
\end{equation}
where $\alpha_{\mathrm{CE}}$ represents the fraction of orbital energy that is
used to unbind the envelope usually called the common envelope efficiency
\citep{paczynski76}. 

The binding energy of the envelope is often assumed to be equal to the
gravitational energy of the envelope: 
\begin{equation}\label{eq:Egr}
E_\mathrm{bind} = E_\mathrm{gr}=-\frac{G M_\mathrm{1} M_\mathrm{1,e}}{\lambda R_\mathrm{1}},
\end{equation}
where $M_\mathrm{1}$, $M_\mathrm{1,e}$ and $R_\mathrm{1}$ are the total mass,
envelope mass and radius of the primary star, and $\lambda$ is a binding
energy parameter. Although very often ignored, the binding energy parameter
strongly depends on the mass and evolutionary state of the white dwarf
progenitor when filling its Roche-lobe. This is especially true if the
recombination energy $U_\mathrm{rec}$ available within the envelope supports
the ejection process. Therefore, a more general form for the binding energy
equation is: 
\begin{equation}\label{eq:Eball}
E_\mathrm{bind}=\int_{M_\mathrm{1,c}}^{M_\mathrm{1}}-\frac{G m}{r(m)}dm + \alpha_{\mathrm{rec}}\int_{M_\mathrm{1,c}}^{M_\mathrm{1}}U_\mathrm{rec}(m),
\end{equation}
where $\alpha_\mathrm{rec}$ is the fraction of recombination energy that
contributes to the ejection process. It is of outstanding importance for our
understanding of compact binary star evolution to observationally constrain
the values of both common envelope efficiencies and to investigate possible
dependencies on the binary parameters. 

\begin{table}
 \centering
  \caption{Physical and binary parameters of \targ.}
  \label{tab:params}
  \begin{tabular}{@{}lc@{}}
  \hline
  Parameter                          & Value \\
  \hline
  RA                                 & 15:02:22.4896 \\
  Dec                                & -29:41:15.666 \\
  $B$                                & 11.578 \\
  $V$                                & 11.245 \\
  $R$                                & 11.030 \\
  $J$                                & 10.167 \\
  $H$                                & 9.918 \\
  $K$                                & 9.853 \\
  Distance (pc)                      & 250--320 \\
  Orbital period (days)              & $0.498688(26)$ \\
  $K_\mathrm{MS}$ (\kms)               & $65.0\pm0.3$ \\
  $v_\mathrm{rot}\sin{i}$ (\kms)       & $75.0\pm3.0$ \\
  Inclination ($^\circ$)              & 33--43 \\
  Separation (\RSUN)                 & 3.20--3.28 \\
  White dwarf mass (\MSUN)           & 0.52--0.67 \\
  White dwarf temperature (K)        & 19,500--21,000 \\
  Main-sequence star spectral type   & F8 \\
  Main-sequence star temperature (K) & 6300--6500 \\
  Main-sequence star surface gravity & 4.31--4.48 \\
  Main-sequence star mass (\MSUN)    & 1.22--1.25 \\
  Main-sequence star radius (\RSUN)  & 1.18--1.40 \\
  \hline
  \end{tabular}
\end{table}

With its period of less than half a day {\targ} is the first short orbital
period ($P_{\mathrm{orb}}<1$\,day) post common envelope binary (PCEB) with a
massive secondary star (spectral type earlier than K). The two systems that
have been previously discovered are IK\,Peg and KOI-3278 with much longer
orbital periods of 21.722 and 88.18 days respectively
\citep{wonnacott93,kruse14}. These two systems would have formed through
common envelope evolution only if the common envelope efficiency
$\alpha_{\mathrm{CE}}$ has been larger than it seems to be required to
understand PCEBs with M dwarf secondaries, where
$\alpha_{\mathrm{CE}}\sim\,0.25-0.5$ seems to work best 
\citep{zorotovic10,rebassa12,toonen13,camacho14} and/or if in addition to a
high fraction of the released orbital energy also recombination energy
contributed to some degree to the ejection process expelling the envelope
\citep{zorotovic10,zorotovic14}. 

To investigate whether {\targ} confirms or disproves the trend of larger
efficiencies being required to understand PCEBs with massive secondaries we
follow \citet{zorotovic10} and reconstruct the evolutionary history of 
the system using the binary star evolution code BSE \citep{hurley02} and the
parameter constraints derived in the previous sections, i.e. $M_\mathrm{WD}
= 0.52-0.67$\MSUN, $M_\mathrm{MS} = 1.22-1.25$\MSUN, P$_\mathrm{orb} =
0.498688$ days and T$_\mathrm{eff,WD} = 19500-21000$K. The current cooling
age of the white dwarf is only $\sim\,0.1$\,Gyr \citep{fontaine01} hence the
current orbital period is virtually identical to the period the system had at
the end of the common envelope (even if efficient magnetic braking is
assumed). Reconstructing the common envelope evolution as in
\citet{zorotovic14} we find a large range of possible values for the common
envelope efficiencies for {\targ} and different possible evolutionary stages
at which the progenitor of the white dwarf could have filled its Roche-lobe. 

\begin{figure}
\begin{center}
 \includegraphics[width=0.98\columnwidth]{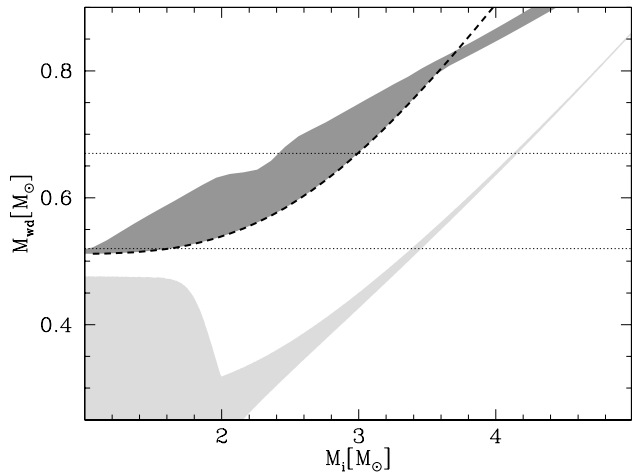}
 \caption{The mass of the white dwarf after the common envelope phase as a
   function of its initial progenitor mass. The resultant white dwarf mass
   depends heavily upon the evolutionary stage at which the common envelope
   started. We show the results for a progenitor on the first giant branch
   (light grey), on the early AGB (dashed line) and on the TP-AGB (dark
   grey). The horizontal lines correspond to the range of white dwarf masses
   in agreement with our observations.}
 \label{fig:rec}
 \end{center}
\end{figure}

Figure~\ref{fig:rec} shows the solutions with $\alpha_{\mathrm{CE}}<1$ in a
final versus initial mass plot for the primary if contributions from
recombination energy are ignored (i.e. $\alpha_{\mathrm{rec}}=0.0$). Solutions
exist for massive progenitors on the first giant branch (FGB), on the early
asymptotic giant branch (AGB) and on the thermally pulsating-AGB
  (TP-AGB). The more massive the progenitor, the larger the value of the
common envelope efficiency must have been and the younger is the system
today. The possible ranges are $\alpha=0.90-1.0$ for a $3.40-4.13${\MSUN}
progenitor that fills its Roche-lobe on the FGB (becoming first a naked
  helium star then eventually a carbon-oxygen white dwarf,
  \citealt{hurley02}) at an age of $0.17-0.27$\,Gyr; $\alpha = 0.08-0.89$ if
the progenitor reached the early AGB, after $0.48-2.49$ Gyr and with an
initial mass of $1.61-2.99$\MSUN; and finally $\alpha_{\mathrm{CE}}=0.05-0.20$
for a $1.37-2.99$\MSUN progenitor that filled its Roche-lobe on the TP-AGB
at an age of $0.48-4.15$\,Gyr, making it the oldest option. In the latter case
we get the largest ranges as the core mass on the TP-AGB reaches the
current white dwarf mass for a large range of initial masses. We therefore
consider this the most likely scenario. This is further supported by our
measurement of the radius of the secondary. Assuming solar metallicity a
$1.235${\MSUN} main sequence star needs $\sim1.8$\,Gyr to expand from its ZAMS
radius to the current radius of the secondary of $1.35$\RSUN. We therefore
conclude that the most consistent scenario for the evolutionary history of
{\targ} is that the progenitor of the white dwarf was of a relatively low mass
($\sim2$\,\MSUN) and filled its Roche-lobe on the TP-AGB, which implies
that the common envelope efficiency must have been small
$\alpha_{\mathrm{CE}}=0.05-0.20$. This is much smaller than values obtained
for the long orbital period systems with secondary stars of similar mass
(IK\,Peg and KOI-3278). In addition, contributions from recombination energy
are not required to understand the existence of \targ. If recombination energy
contributed to expelling the envelope of the progenitor of the white dwarf,
the fraction of orbital energy lost during the common envelope is reduced even
further while the possible initial masses would remain virtually identical. It
therefore seems that a simple dependence of the total common envelope
efficiency on the secondary mass as speculated by \citet{zorotovic14} remains
an incomplete prescription for common envelope evolution and other parameters
such as perhaps the evolutionary state of the progenitor when it fills its
Roche-lobe may play an important role. Characterizing more PCEBs with massive
secondaries is therefore urgently required to progress with our understanding
of compact binary evolution which is directly related to our understanding of
SN\,Ia progenitor systems.  

\subsection{The future of \targ}

While it is clear that PCEBs with M dwarf secondaries evolve into cataclysmic
variables (CVs) as long as the mass transfer will be dynamically stable, the
future of PCEBs with more massive secondary stars is more uncertain and
potentially very interesting as the total mass of the binary usually exceeds
the Chandrasehkar limit. Indeed, PCEBs with massive (G or F type) secondaries
are the progenitors of SN\,Ia explosions for both the double and the single
degenerate channel. If the separation of a given PCEB is large it potentially
survives a second common envelope forming a double degenerate system which may
then merge and produce a SN\,Ia. If on the other hand, the system is close it
might start thermal time scale mass transfer and reach the rates required for
stable nuclear burning on the surface of the white dwarf allowing its mass to
increase and potentially reach the Chandrasehkar limit. 

\begin{figure}
\begin{center}
 \vspace{3mm}
 \includegraphics[width=1.05\columnwidth,bb=10 0 576 380]{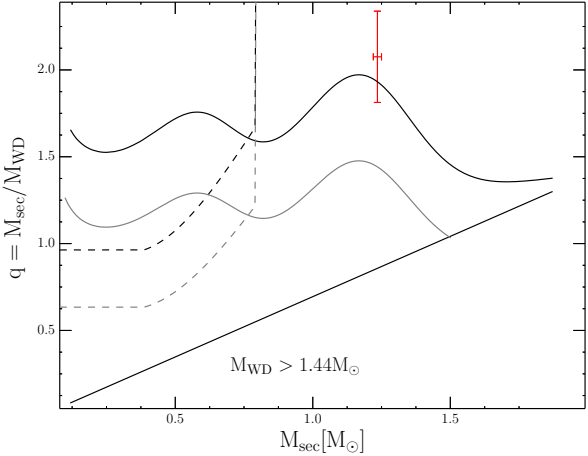}
 \caption{Critical mass ratio for thermal time scale mass transfer as a
   function of the main-sequence star's mass for conservative (grey) and
   non-conservative (black) mass transfer. The position of {\targ} is
   indicated in red. For completeness we also provide the limits for dynamical
   time scale mass transfer (dashed lines) where we used the fits to detailed
   calculations of the adiabatic mass-radius exponent \citep{hjellming89}.} 
 \label{fig:stab}
 \end{center}
\end{figure}

With its short orbital period and given that the system is rather young,
{\targ} will clearly start mass transfer before the secondary evolves
off the main-sequence. It is also clear that the system will not undergo
dynamical time scale mass transfer as the critical mass ratio for dynamically
unstable mass transfer is $q_{\mathrm{cr}}\gae2.5$ \citep{ge13} for donor
stars with masses and radii similar to the Sun. It is less clear, however, if
the mass transfer will be thermally stable (i.e. stable against thermal
  time scale mass transfer). The critical mass ratio for
marginal stability against thermal time scale mass transfer in a semi-detached
binary star can be obtained by equating the mass-radius exponent of the 
Roche-lobe and the star, i.e. 
\begin{equation}
\zeta_{\mathrm{th}}=\frac{d\ln(R_\mathrm{MS})}{d\ln(M_\mathrm{MS})}=
\frac{d\ln(R_{\mathrm{L}})}{d\ln(M_\mathrm{MS})}, 
\end{equation}
\citep[see also e.g.][]{dekool92}, where the Roche-lobe radius
($R_{\mathrm{L}}$) is a function of the mass ratio and the binary
separation. Despite the secondary in {\targ} being slightly evolved we
  can get a first hint about the future of the system by calculating the limit
  implied by Eq.\,8 using the ZAMS M-R relation from \citet{tout96}. For
  conservative mass transfer we get a critical mass ratio for
thermal time scale mass transfer of $q_{\mathrm{cr}}=1.4-1.48$ for \targ. Thus
in the case of conservative mass transfer the system will certainly experience
thermal time scale mass transfer. 

However, the assumption of conservative mass transfer is not necessarily
correct. As long as the mass transfer rate stays below the critical value for
stable hydrogen burning on the white dwarf, mass transfer will most likely not
be conservative. Instead, nova eruptions will occur, the white dwarf mass
will remain nearly constant and angular momentum will be taken away from the
system by the expelled material. In Figure~\ref{fig:stab} we show the critical
values for the mass ratio as a function of secondary mass for both
conservative  and non-conservative mass transfer. In the latter case we assume
the mass expelled during nova eruptions to carry the specific angular momentum
of the white dwarf. The white dwarf mass is assumed to be constant and
  the mass (and angular momentum) loss to be continuous which has been shown
  to be equivalent to a discontinuous sequence of nova cycles in terms of the
  secular evolution of CVs \citep{schenker98}. These assumptions are typical
assumptions for CVs \citep[e.g.][]{ritter88}. The position of {\targ} is very
close to the limit for thermal time scale mass transfer in the case of
non-conservative mass transfer so the system may indeed be a progenitor of a
super soft source.

\begin{figure}
\begin{center}
 \includegraphics[width=\columnwidth,clip,bb=12 8 530 390]{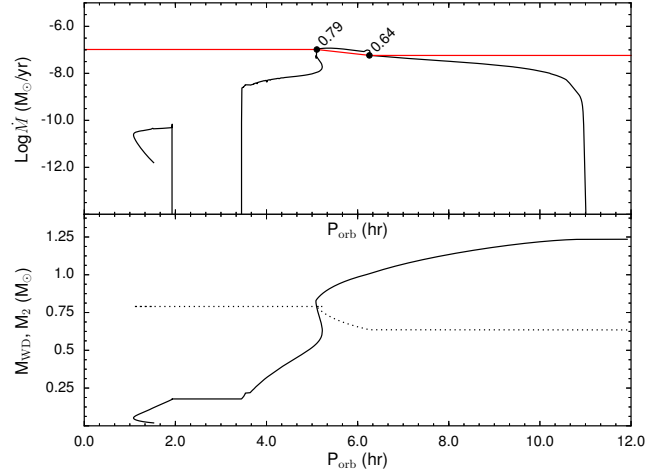}
 \caption{The predicted future evolution of the mass transfer rate (top
   panel), the white dwarf mass (dotted line, bottom panel), and the secondary
   star mass (solid line, bottom panel) as a function of orbital period. The
   horizontal red line in the top panel represents the mass transfer rate
   required for stable hydrogen burning for the respective mass of the white
   dwarf (i.e. when the mass transfer rate is above this line stable hydrogen
   burning occurs). At $P_{\mathrm{orb}}\sim6.5$\,h {\targ} reaches this
   critical mass transfer rate. As a consequence mass transfer becomes
   conservative and the white dwarf grows in mass. At
   $P_{\mathrm{orb}}\sim5$\,h the white dwarf mass has reached $0.79$\MSUN 
   and the mass transfer rate drops below the rate required for stable
   burning. At this moment {\targ} becomes a ``normal'' CV with
   non-conservative stable mass transfer driven by angular momentum loss only
   and a constant white dwarf mass.}
 \label{fig:CV}
 \end{center}
\end{figure}

In order to investigate the future of {\targ} in more detail we performed
dedicated simulations using MESA \citep{paxton11}. We assume the initial
masses of the two stars to be $M_\mathrm{MS}=1.235${\MSUN} and
$M_{\mathrm{WD}}=0.63${\MSUN} and secondary radius of $1.35$\RSUN, which
corresponds to the mean values of the ranges  in agreement with our
observations. The starting model was obtained by evolving the secondary until
it reaches the required radius (after $1.8$\,Gyr) and the white dwarf is
approximated as a point mass. We assume mass transfer to be nearly
conservative (90 per cent of the transferred mass remains on the white dwarf)
if the mass transfer is large enough to generate stable hydrogen burning on the
surface of the white dwarf as we run into numerical problems in the fully
conservative case. For mass transfer rates below the limit for stable burning
we assume that all the accreted mass is expelled during nova eruptions leaving
the system with the specific angular momentum of the white dwarf. Finally, for
angular momentum loss due to magnetic braking we assume the standard
prescription from \citet{rappaport83}.  

Based on these assumptions, the future of {\targ} can be estimated. The black
line in the top panel of Figure~\ref{fig:CV} shows the predicted mass transfer
rate of {\targ} as a function of orbital period. At an orbital 
period of 11.9\,h (in less than 5\,Myr) mass transfer will start. The mass
transfer rate will quickly increase as the system is thermally unstable, but
does not reach the limit for stable hydrogen burning until the system has an
orbital period of 6.2\,h (13.1\,Myr from now). At this point, mass transfer
becomes conservative (we assume that 90 per cent of the burned material
remains on the white dwarf). As a consequence, the mass transfer rate
increases significantly and the white dwarf mass grows (bottom panel). When
the white dwarf mass reaches $0.79${\MSUN} (14.1\,Myr from now) the mass
ratio will be close to unity (secondary mass of $0.8${\MSUN} at an orbital
period of 5.1\,h) and the system becomes stable against (conservative)
thermal time scale mass transfer. Therefore the accretion rate drops below the
value required for stable hydrogen burning. From this moment on, {\targ} will
behave as expected for normal CVs: after 105\,Myr it will enter the period
gap and restart mass transfer at a much lower rate at an orbital period of
2.1\,h (1.5\,Gyr from now). Finally, the system will reach the orbital
period minimum ($\sim$1.1\,h) 4.8\,Gyr from now. 

\begin{figure}
\begin{center}
 \includegraphics[width=0.97\columnwidth,bb=20 7 530 390]{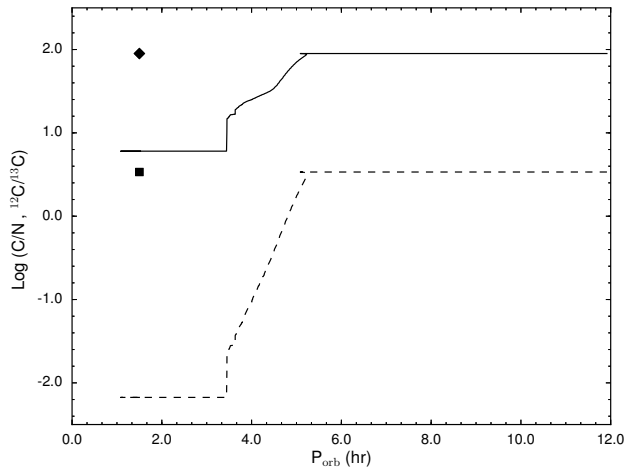}
 \caption{Surface abundance ratios (C/N, dashed line and C$_{12}/$C$_{13}$,
   solid line) for the main-sequence star in {\targ} as a function of orbital
   period. Once a deep convective envelope forms (P$_{\mathrm{orb}}\sim$5\,h)
   both abundance ratios substantially decrease as carbon depleted material is
   dredged up from the inner regions of the star where CNO burning has taken
   place. In contrast, the abundance ratios of a CV born with a low-mass donor
   (which never began CNO burning) will remain largely fixed, even when the
   star becomes fully convective. We show the abundance ratios for such a
   ``normal'' CV (C/N, black square, C$_{12}/$C$_{13}$, black diamond).}  
 \label{fig:abun}
 \end{center}
\end{figure}

While the predicted final evolution of the period and mass transfer rate of
{\targ} is identical to those of CVs descending from PCEBs with less massive
secondaries, its past as a PCEB containing a more massive secondary star will 
be imprinted in the relative abundances of carbon and nitrogen of the accreted
material as soon as the outer convection zone reaches regions containing CNO
processed material. Decreased values for C/N and C$_{12}/$C$_{13}$ in CVs
descending from thermal time scale mass transfer systems have been predicted
by \citet{schenker02} and found in CVs based on the presence of C/N line
ratios measured with HST by \citet{gaensicke03}. In Figure~\ref{fig:abun} we
predict the C/N and C$_{12}/$C$_{13}$ abundance ratios for {\targ} as a function
of orbital period. As soon as the secondary star starts to develop a deep
convective envelope both surface abundance ratios decrease
significantly. Most dramatically, C/N decreases by several orders of
magnitude. {\targ} is therefore likely a progenitor of the CVs with decreased
C/N line ratios found by \citet{gaensicke03} which are frequently called
{\em failed} SN\,Ia.

A final note concerning the future of {\targ} concerns the expected white
dwarf mass when the systems becomes a CV. The mean white dwarf mass in CVs
derived from observations ($\sim0.8$\MSUN, \citealt{zorotovic11}) is
significantly higher than predicted by binary population models of CVs and
significantly larger than the mean value observed in CV progenitors containing
low mass secondaries \citep[see][for details]{zorotovic11}. The value
predicted by our simulations for {\targ}, however, is very close to the mean
white dwarf mass of the observed CV sample. One could therefore speculate that
if the number of CVs descending from PCEBs with massive secondaries has been
underestimated previously, the problem with the white dwarf masses in CVs
could be solved.  Assuming that a large number of CVs are descendants from
PCEBS with massive secondary stars such as {\targ}  indeed predicts an
increased number of CVs with white dwarf masses $\sim0.8$\MSUN. However, the
large number of CVs with evolved secondaries  predicted in this scenario
violates the general explanation for the orbital period gap \citep[see][for
more details]{wjinen15}. We therefore believe that the fraction of CVs
descending from systems with initially massive secondaries does not
significantly exceed the $\sim10$ per cent of CVs with decreased C/N 
line ratios as measured by \citet{gaensicke03} and the white dwarf mass
problem remains to be solved. 
 
We emphasize that the above evolutionary scenario for {\targ} is based on
several assumptions such as the strength of magnetic braking, the angular
momentum taken away from the system during nova eruptions, and the fraction of
mass that is lost during thermal time scale mass transfer. However, our
simulation shows that it is likely that  {\targ} is the first known progenitor
of a super-soft source.  

\section{Conclusions}

We have identified {\targ} as a close binary consisting of a white dwarf and
an F8 star that is extremely close to filling its Roche lobe. Assuming
  that the F star is tidally locked, we constrain the masses of the two stars
to $M_\mathrm{WD}= 0.52-0.67$\MSUN, $M_\mathrm{MS} = 1.22-1.25${\MSUN} and the
radius of the F star to $R_\mathrm{MS} = 1.18-1.40$\RSUN. The white dwarf is
still hot ($19,500-21,000$K) and so the system only emerged from the common
envelope 0.1\,Gyr ago. Its progenitor was likely on the TP-AGB when the
common envelope began. However, this means that the common envelope efficiency
was quite low, in contrast to other systems containing white dwarfs with early
type companions, implying that the efficiency depends upon more than just the
mass of the companion star.

{\targ} will become a semi-detached system in less than 5\,Myr and will
undergo a short phase ($\sim1$\,Myr) of thermal timescale mass-transfer in
13.1\,Myr resulting in a roughly 20 per cent increase in the white dwarf's
mass, still well short of the Chandrasekhar limit, after which it will become
a standard cataclysmic variable system.

Although we have placed some constraints on the binary parameters, these will
be greatly improved with an accurate distance measurement from Gaia which
should reach a precision of $\sim0.3\%$ at the expected distance of
{\targ}\footnote{http://www.cosmos.esa.int/web/gaia/science-performance}. This will dramatically reduce the reliance on evolutionary models in
our measurements allowing us to more rigorously test the evolutionary
scenarios of this (and similar) systems.

\section*{Acknowledgments}

We thank the referee for useful comments and suggestions. SGP, MZ and CT
acknowledge financial support from FONDECYT in the form of grant numbers
3140585, 3130559 and 1120338. AB acknowledges financial support from  Proyecto
FONDECYT de Iniciaci{\'o}n 11140572. The research leading to these results has
received funding from the European Research Council under the European Union's
Seventh Framework Programme (FP/2007-2013) / ERC Grant Agreement n. 320964
(WDTracer). MRS thanks for support from FONDECYT (1141269) and Millennium
Science Initiative, Chilean ministry of Economy: Nucleus RC130007. ARM
acknowledges   financial  support   from   the  Postdoctoral Science
Foundation of China  (grants 2013M530470 and 2014T70010)  and from the
Research  Fund  for International  Young  Scientists  by the  National Natural 
Science   Foundation  of   China  (grant   11350110496). RB is supported by
CONICYT-PCHA/Doctorado Nacional. RB acknowledge additional support from
project IC120009 "Millenium Institute of Astrophysics (MAS)" of the Millennium
Science Initiative, Chilean Ministry of Economy. AJ acknowledges support from
the Ministry for the Economy, Development, and Tourism's Programa Iniciativa
Cient{\'i}fica Milenio through grant IC 120009, awarded to the Millennium
Institute of Astrophysics (MAS),  FONDECYT  project 1130857 and from BASAL
CATA PFB-06. Based on observations made with the NASA/ESA Hubble Space
Telescope, obtained  at the Space Telescope Science Institute, which is
operated by the Association of Universities for Research in Astronomy, Inc.,
under NASA contract NAS 5-26555. These observations are associated with
program \#13704. The results presented in this paper are based on observations
collected at the European Southern Observatory under programme ID
094.D-0344. The FEROS observations were obtained via Max Planck Institute for
Astronomy (MPIA) guaranteed time. This publication makes use of VOSA,
developed under the Spanish Virtual Observatory project supported from the
Spanish MICINN through grant AyA2008-02156. 

\bibliographystyle{mn_new}
\bibliography{tyc6760_v3}

\begin{thebibliography}{58}
\expandafter\ifx\csname natexlab\endcsname\relax\def\natexlab#1{#1}\fi

\bibitem[{{Allard} et~al.(2012){Allard}, {Homeier}, \& {Freytag}}]{allard12}
{Allard}, F., {Homeier}, D., {Freytag}, B., 2012, Royal Society of London
  Philosophical Transactions Series A, 370, 2765

\bibitem[{{Bayo} et~al.(2008){Bayo}, {Rodrigo}, {Barrado Y Navascu{\'e}s},
  {Solano}, {Guti{\'e}rrez}, {Morales-Calder{\'o}n}, \& {Allard}}]{bayo08}
{Bayo}, A., {Rodrigo}, C., {Barrado Y Navascu{\'e}s}, D., {Solano}, E.,
  {Guti{\'e}rrez}, R., {Morales-Calder{\'o}n}, M., {Allard}, F., 2008, \aap,
  492, 277

\bibitem[{{Branch} \& {Tammann}(1992)}]{branch92}
{Branch}, D., {Tammann}, G.~A., 1992, \araa, 30, 359

\bibitem[{{Bressan} et~al.(2012){Bressan}, {Marigo}, {Girardi}, {Salasnich},
  {Dal Cero}, {Rubele}, \& {Nanni}}]{bressan12}
{Bressan}, A., {Marigo}, P., {Girardi}, L., {Salasnich}, B., {Dal Cero}, C.,
  {Rubele}, S., {Nanni}, A., 2012, \mnras, 427, 127

\bibitem[{{Burleigh} et~al.(1997){Burleigh}, {Barstow}, \&
  {Fleming}}]{burleigh97}
{Burleigh}, M.~R., {Barstow}, M.~A., {Fleming}, T.~A., 1997, \mnras, 287, 381

\bibitem[{{Camacho} et~al.(2014){Camacho}, {Torres}, {Garc{\'{\i}}a-Berro},
  {Zorotovic}, {Schreiber}, {Rebassa-Mansergas}, {Nebot G{\'o}mez-Mor{\'a}n},
  \& {G{\"a}nsicke}}]{camacho14}
{Camacho}, J., {Torres}, S., {Garc{\'{\i}}a-Berro}, E., {Zorotovic}, M.,
  {Schreiber}, M.~R., {Rebassa-Mansergas}, A., {Nebot G{\'o}mez-Mor{\'a}n}, A.,
  {G{\"a}nsicke}, B.~T., 2014, \aap, 566, A86

\bibitem[{{Claret}(2004)}]{claret04}
{Claret}, A., 2004, \aap, 428, 1001

\bibitem[{{Coelho} et~al.(2005){Coelho}, {Barbuy}, {Mel{\'e}ndez}, {Schiavon},
  \& {Castilho}}]{coelho05}
{Coelho}, P., {Barbuy}, B., {Mel{\'e}ndez}, J., {Schiavon}, R.~P., {Castilho},
  B.~V., 2005, \aap, 443, 735

\bibitem[{{de Kool}(1992)}]{dekool92}
{de Kool}, M., 1992, \aap, 261, 188

\bibitem[{{D'Odorico} et~al.(2006)}]{dodorico06}
{D'Odorico}, S., et~al., 2006, in Proc. SPIE, vol. 6269, p.~98

\bibitem[{{Fink} et~al.(2007){Fink}, {Hillebrandt}, \& {R{\"o}pke}}]{fink07}
{Fink}, M., {Hillebrandt}, W., {R{\"o}pke}, F.~K., 2007, \aap, 476, 1133

\bibitem[{{Fink} et~al.(2010){Fink}, {R{\"o}pke}, {Hillebrandt}, {Seitenzahl},
  {Sim}, \& {Kromer}}]{fink10}
{Fink}, M., {R{\"o}pke}, F.~K., {Hillebrandt}, W., {Seitenzahl}, I.~R., {Sim},
  S.~A., {Kromer}, M., 2010, \aap, 514, A53

\bibitem[{{Fontaine} et~al.(2001){Fontaine}, {Brassard}, \&
  {Bergeron}}]{fontaine01}
{Fontaine}, G., {Brassard}, P., {Bergeron}, P., 2001, \pasp, 113, 409

\bibitem[{{G{\"a}nsicke} et~al.(2003)}]{gaensicke03}
{G{\"a}nsicke}, B.~T., et~al., 2003, \apj, 594, 443

\bibitem[{{Ge} et~al.(2013){Ge}, {Webbink}, {Chen}, \& {Han}}]{ge13}
{Ge}, H., {Webbink}, R.~F., {Chen}, X., {Han}, Z., 2013, in {Zhang}, C.~M.,
  {Belloni}, T., {M{\'e}ndez}, M., {Zhang}, S.~N., eds., IAU Symposium, vol.
  290 of \emph{IAU Symposium}, p. 213

\bibitem[{{Greiner}(2000)}]{greiner00}
{Greiner}, J., 2000, \na, 5, 137

\bibitem[{{Hjellming}(1989)}]{hjellming89}
{Hjellming}, M.~S., 1989, {Rapid mass transfer in binary systems}, Ph.D.
  thesis, Illinois Univ.~at Urbana-Champaign, Savoy.

\bibitem[{{Holberg} et~al.(2013){Holberg}, {Oswalt}, {Sion}, {Barstow}, \&
  {Burleigh}}]{holberg13}
{Holberg}, J.~B., {Oswalt}, T.~D., {Sion}, E.~M., {Barstow}, M.~A., {Burleigh},
  M.~R., 2013, \mnras, 435, 2077

\bibitem[{{Hubeny} \& {Lanz}(1995)}]{hubeny95}
{Hubeny}, I., {Lanz}, T., 1995, \apj, 439, 875

\bibitem[{{Hurley} et~al.(2002){Hurley}, {Tout}, \& {Pols}}]{hurley02}
{Hurley}, J.~R., {Tout}, C.~A., {Pols}, O.~R., 2002, \mnras, 329, 897

\bibitem[{{Husser} et~al.(2013){Husser}, {Wende-von Berg}, {Dreizler},
  {Homeier}, {Reiners}, {Barman}, \& {Hauschildt}}]{husser13}
{Husser}, T.-O., {Wende-von Berg}, S., {Dreizler}, S., {Homeier}, D.,
  {Reiners}, A., {Barman}, T., {Hauschildt}, P.~H., 2013, \aap, 553, A6

\bibitem[{{Jord{\'a}n} et~al.(2014)}]{jordan14}
{Jord{\'a}n}, A., et~al., 2014, \aj, 148, 29

\bibitem[{{Kordopatis} et~al.(2013)}]{kordopatis13}
{Kordopatis}, G., et~al., 2013, \aj, 146, 134

\bibitem[{{Kruse} \& {Agol}(2014)}]{kruse14}
{Kruse}, E., {Agol}, E., 2014, Science, 344, 275

\bibitem[{{Marsh}(1989)}]{marsh89}
{Marsh}, T.~R., 1989, \pasp, 101, 1032

\bibitem[{{Martin} et~al.(2005)}]{martin05}
{Martin}, D.~C., et~al., 2005, \apjl, 619, L1

\bibitem[{{Maxted} et~al.(2009){Maxted}, {G{\"a}nsicke}, {Burleigh},
  {Southworth}, {Marsh}, {Napiwotzki}, {Nelemans}, \& {Wood}}]{maxted09}
{Maxted}, P.~F.~L., {G{\"a}nsicke}, B.~T., {Burleigh}, M.~R., {Southworth}, J.,
  {Marsh}, T.~R., {Napiwotzki}, R., {Nelemans}, G., {Wood}, P.~L., 2009,
  \mnras, 400, 2012

\bibitem[{{Napiwotzki} et~al.(2003)}]{napiwotzki03}
{Napiwotzki}, R., et~al., 2003, The Messenger, 112, 25

\bibitem[{{Paczynski}(1976)}]{paczynski76}
{Paczynski}, B., 1976, in {Eggleton}, P., {Mitton}, S., {Whelan}, J., eds.,
  Structure and Evolution of Close Binary Systems, vol.~73 of \emph{IAU
  Symposium}, p.~75

\bibitem[{{Parsons} et~al.(2012)}]{parsons12}
{Parsons}, S.~G., et~al., 2012, \mnras, 420, 3281

\bibitem[{{Paxton} et~al.(2011){Paxton}, {Bildsten}, {Dotter}, {Herwig},
  {Lesaffre}, \& {Timmes}}]{paxton11}
{Paxton}, B., {Bildsten}, L., {Dotter}, A., {Herwig}, F., {Lesaffre}, P.,
  {Timmes}, F., 2011, \apjs, 192, 3

\bibitem[{{Perlmutter} et~al.(1999)}]{perlmutter99}
{Perlmutter}, S., et~al., 1999, \apj, 517, 565

\bibitem[{{Pyrzas} et~al.(2012)}]{pyrzas12}
{Pyrzas}, S., et~al., 2012, \mnras, 419, 817

\bibitem[{{Rappaport} et~al.(1983){Rappaport}, {Verbunt}, \&
  {Joss}}]{rappaport83}
{Rappaport}, S., {Verbunt}, F., {Joss}, P.~C., 1983, \apj, 275, 713

\bibitem[{{Rebassa-Mansergas} et~al.(2012{\natexlab{a}}){Rebassa-Mansergas},
  {Nebot G{\'o}mez-Mor{\'a}n}, {Schreiber}, {G{\"a}nsicke}, {Schwope},
  {Gallardo}, \& {Koester}}]{rebassa12b}
{Rebassa-Mansergas}, A., {Nebot G{\'o}mez-Mor{\'a}n}, A., {Schreiber}, M.~R.,
  {G{\"a}nsicke}, B.~T., {Schwope}, A., {Gallardo}, J., {Koester}, D.,
  2012{\natexlab{a}}, \mnras, 419, 806

\bibitem[{{Rebassa-Mansergas} et~al.(2012{\natexlab{b}})}]{rebassa12}
{Rebassa-Mansergas}, A., et~al., 2012{\natexlab{b}}, \mnras, 423, 320

\bibitem[{{Ribeiro} et~al.(2013){Ribeiro}, {Baptista}, {Kafka}, {Dufour},
  {Gianninas}, \& {Fontaine}}]{ribeiro13}
{Ribeiro}, T., {Baptista}, R., {Kafka}, S., {Dufour}, P., {Gianninas}, A.,
  {Fontaine}, G., 2013, \aap, 556, A34

\bibitem[{{Riess} et~al.(1998)}]{riess98}
{Riess}, A.~G., et~al., 1998, \aj, 116, 1009

\bibitem[{{Ritter}(1988)}]{ritter88}
{Ritter}, H., 1988, \aap, 202, 93

\bibitem[{{Schenker} et~al.(1998){Schenker}, {Kolb}, \& {Ritter}}]{schenker98}
{Schenker}, K., {Kolb}, U., {Ritter}, H., 1998, \mnras, 297, 633

\bibitem[{{Schenker} et~al.(2002){Schenker}, {King}, {Kolb}, {Wynn}, \&
  {Zhang}}]{schenker02}
{Schenker}, K., {King}, A.~R., {Kolb}, U., {Wynn}, G.~A., {Zhang}, Z., 2002,
  \mnras, 337, 1105

\bibitem[{{Schlafly} \& {Finkbeiner}(2011)}]{schlafly11}
{Schlafly}, E.~F., {Finkbeiner}, D.~P., 2011, \apj, 737, 103

\bibitem[{{Shen} \& {Moore}(2014)}]{shen14}
{Shen}, K.~J., {Moore}, K., 2014, \apj, 797, 46

\bibitem[{{Skrutskie} et~al.(2006)}]{skrutskie06}
{Skrutskie}, M.~F., et~al., 2006, \aj, 131, 1163

\bibitem[{{Tappert} et~al.(2011){Tappert}, {G{\"a}nsicke}, {Schmidtobreick}, \&
  {Ribeiro}}]{tappert11}
{Tappert}, C., {G{\"a}nsicke}, B.~T., {Schmidtobreick}, L., {Ribeiro}, T.,
  2011, \aap, 532, A129

\bibitem[{{Toonen} \& {Nelemans}(2013)}]{toonen13}
{Toonen}, S., {Nelemans}, G., 2013, \aap, 557, A87

\bibitem[{{Torres} et~al.(2010){Torres}, {Andersen}, \&
  {Gim{\'e}nez}}]{Torres10}
{Torres}, G., {Andersen}, J., {Gim{\'e}nez}, A., 2010, \aapr, 18, 67

\bibitem[{{Tout} et~al.(1996){Tout}, {Pols}, {Eggleton}, \& {Han}}]{tout96}
{Tout}, C.~A., {Pols}, O.~R., {Eggleton}, P.~P., {Han}, Z., 1996, \mnras, 281,
  257

\bibitem[{{Tutukov} \& {Yungelson}(1996)}]{tutukov96}
{Tutukov}, A., {Yungelson}, L., 1996, \mnras, 280, 1035

\bibitem[{{Webbink}(1984)}]{webbink84}
{Webbink}, R.~F., 1984, \apj, 277, 355

\bibitem[{{Whelan} \& {Iben}(1973)}]{whelan73}
{Whelan}, J., {Iben}, Jr., I., 1973, \apj, 186, 1007

\bibitem[{{Wijnen} et~al.(2015){Wijnen}, {Zorotovic}, \&
  {Schreiber}}]{wjinen15}
{Wijnen}, T.~P.~G., {Zorotovic}, M., {Schreiber}, M.~R., 2015, ArXiv e-prints

\bibitem[{{Wonnacott} et~al.(1993){Wonnacott}, {Kellett}, \&
  {Stickland}}]{wonnacott93}
{Wonnacott}, D., {Kellett}, B.~J., {Stickland}, D.~J., 1993, \mnras, 262, 277

\bibitem[{{Wright} et~al.(2010)}]{wright10}
{Wright}, E.~L., et~al., 2010, \aj, 140, 1868

\bibitem[{{Zahn}(1977)}]{zahn77}
{Zahn}, J.-P., 1977, \aap, 57, 383

\bibitem[{{Zorotovic} et~al.(2010){Zorotovic}, {Schreiber}, {G{\"a}nsicke}, \&
  {Nebot G{\'o}mez-Mor{\'a}n}}]{zorotovic10}
{Zorotovic}, M., {Schreiber}, M.~R., {G{\"a}nsicke}, B.~T., {Nebot
  G{\'o}mez-Mor{\'a}n}, A., 2010, \aap, 520, A86

\bibitem[{{Zorotovic} et~al.(2011){Zorotovic}, {Schreiber}, \&
  {G{\"a}nsicke}}]{zorotovic11}
{Zorotovic}, M., {Schreiber}, M.~R., {G{\"a}nsicke}, B.~T., 2011, \aap, 536,
  A42

\bibitem[{{Zorotovic} et~al.(2014){Zorotovic}, {Schreiber}, \&
  {Parsons}}]{zorotovic14}
{Zorotovic}, M., {Schreiber}, M.~R., {Parsons}, S.~G., 2014, \aap, 568, L9

\end{thebibliography}

\label{lastpage}

\end{document}